# A case for Bragg diffraction by a noncollinear (PT)-symmetric antiferromagnet $Cu_2(MoO_4)(SeO_3)$

S. W. Lovesey[1,2,3] and G. van der Laan[2]

[1] *ISIS Facility, STFC, Didcot, Oxfordshire OX11 0QX, UK*
[2] *Diamond Light Source, Harwell Science and Innovation Campus, Didcot, Oxfordshire OX11 0DE, UK*
[3] *Department of Physics, Oxford University, Oxford OX1 3PU, UK*

**Abstract**
Antiferromagnetic compounds chromium sesquioxide ($Cr_2O_3$) and dioxomolybdenum selenite present linear magnetoelectric effects. Anti-inversion symmetry in the corresponding magnetic crystal classes dictate the makeup of magnetic Bragg diffraction patterns. Copper axial and polar magnetic multipoles contribute to resonant x-ray and magnetic neutron amplitudes in a symmetry informed analysis of monoclinic $Cu_2(MoO_4)(SeO_3)$ presented with a view to steering future diffraction experiments. The compound might be viewed as a low-dimensional quantum magnet on account of its crystal structure and extreme Cu spin (S = 1/2).

## I. INTRODUCTION

Néel [1] proposed antiferromagnetism and J. Samuel Smart [2] recognized that magnetic neutron Bragg diffraction could provide the concrete evidence. To this end, diffraction patterns were collected from powdered MnO that presents magnetic order below a temperature ≈ 122 K. A difference in patterns from samples at room temperature and a temperature ≈ 80 K revealed strong magnetic Bragg spots at positions not allowed on the basis of the chemical face-centred cubic unit cell of MnO. The spots could be indexed on a magnetic unit cell twice as large as the chemical unit cell. Antiferromagnetic domain structure in rutile $MnF_2$ was studied with interference between nuclear and magnetic contributions to scattering amplitudes for polarized neutrons [3]. No such interference is permitted in neutron diffraction by antiferromagnetic chromium sesquioxide ($Cr_2O_3$), for example, because it presents a linear magnetoelectric effect. The magnetic structure of $Cr_2O_3$ was established through neutron [4], x-ray [5], and magnetic-susceptibility [6] measurements. Thereafter, Dzyaloshinskii [7] predicted and Astrov [8] measured a magnetoelectric effect. The effect is allowed when parity (P) and time (T) symmetries are paired in the magnetic crystal class, i.e., (PT)-symmetry is a consequence of anti-inversion ($\bar{1}'$). Other magnetoelectric compounds investigated by Bragg diffraction include $GdB_4$ (magnetic crystal class 4/m'm'm') [9], CuMnAs (m'mm) [10] and $Cu_2(MoO_4)(SeO_3)$ (2'/m) [11], where the latter monoclinic transition metal dioxomolybdenum selenite is the subject of the present investigation. It could be viewed as a low-dimensional quantum magnet on account of its crystal structure and extreme Cu spin (S = 1/2). (PT)-symmetry protects resonant x-ray Bragg diffraction patterns against coupling to helicity (handedness) in the primary beam, which is the counterpart in x-ray diffraction to forbidden interference between nuclear and magnetic amplitudes in polarized neutron diffraction by a magnetoelectric compound [12].

We study some consequences of the monoclinic magnetic structure for $Cu_2(MoO_4)(SeO_3)$ proposed by Piyawongwatthana *et al*. [11]. Specifically, a case for additional experiments using Bragg diffraction is built from symmetry informed magnetic amplitudes for resonant x-ray and neutron diffraction. In the latter, for example, the magnetization distribution includes direct information on the spin-orbit coupling mentioned by Piyawongwatthana *et al*. [11] as a likely source of large departures from the naive value $[2\sqrt{S(S+1)}]$ for effective moments along axes a and b. Since the magnetic structure is (PT)-symmetric the matrix of x-ray amplitudes is purely real and diffraction is impervious to helicity. Copper ions are permitted Dirac multipoles that are polar and magnetic, e.g., a Dirac dipole also known as an anapole. Corresponding x-ray and neutron diffraction amplitudes are presented to guide future experimental studies.

## II. MAGNETIC STRUCTURE

The magnetic structure $P2_1'/c$ (No.14.77, [13]) for $Cu_2(MoO_4)(SeO_3)$ is depicted in Fig. 1 with copper ions in Wyckoff general positions 4e [11]. It is a commensurate structure with propagation vector = (0, 0, 0). Sites 4e are devoid of symmetry. Copper ions support axial (parity even) and polar (Dirac, parity odd) magnetic multipoles. An electronic structure factor (SF) from which all diffraction amplitudes are calculated is the principal subject of an Appendix. It depends explicitly on parity and time-reversal discrete symmetries recorded through two signatures; parity signature $\sigma_\pi = +1\ (-1)$ for parity even (parity odd) and time signature $\sigma_\theta = +1\ (-1)$ for time even (time odd, magnetic). For magnetic neutron scattering (Section IV) all multipoles are time odd and can be different from zero in the magnetically ordered phase. Total scattering amplitudes are a sum of nuclear and magnetic contributions that do not overlap, i.e., the two contributions are 90º apart. However, nuclear contributions are absent in space group (basis) forbidden Bragg spots that we study [11]. Turning to resonant x-ray diffraction (Section III), angular anisotropy in atomic charge distributions allows non-magnetic ($\sigma_\theta = +1$) multipoles to contribute in space group forbidden Bragg spots. Properties of our electronic multipoles for x-ray and neutron diffraction are gathered in Table I [14, 15, 16].

The monoclinic cell possesses unique axis b [11]. Magnetic properties of copper ions (multipoles) are referred to orthogonal axes labelled ($\xi, \eta, \zeta$) derived in standard form from monoclinic axes; specifically, $\xi \propto a^*_m$, $\eta \propto b_m$, $\zeta \propto c_m$.

## III. X-RAY DIFFRACTION

Tuning the energy of the x-rays to an atomic resonance has two obvious benefits in diffraction experiments [14, 15, 17-20]. In the first place, there is a welcome enhancement of Bragg spot intensities and, secondly, spots are element specific. States of x-ray polarization,

Bragg angle θ, and the plane of scattering are shown in Fig. 2. A conventional labelling of linear photon polarization states places **σ** = (0, 0, 1) and **π** = (cos(θ), sin(θ), 0) perpendicular and parallel to the plane of scattering, respectively. Secondary states **σ′** = **σ** and **π′** = (cos(θ), − sin(θ), 0).

The x-ray scattering length in the unrotated channel of polarization σ → σ′, say, is modelled by (σ′σ)/D(E). In this instance, the resonant denominator is replaced by a sharp oscillator D(E) = {[E − Δ + iΓ/2]/Δ} with the x-ray energy E in the near vicinity of an atomic resonance Δ of total width Γ, namely, E ≈ Δ and Γ << Δ. The cited energy-integrated scattering amplitude (σ′σ), one of four amplitudes, is studied using standard tools and methods from atomic physics and crystallography. In the first place, a vast spectrum of virtual intermediate states makes the x-ray scattering length extremely complicated. It can be truncated following closely steps in celebrated studies by Judd and Ofelt of optical absorption intensities of rare-earth ions [21-23, 24]. An intermediate level of truncation used here reproduces sum rules for axial dichroic signals created by electric dipole - electric dipole (E1-E1) or electric quadrupole - electric quadrupole (E2-E2) absorption events [25]. The attendant calculation presented in Ref. [24] and Section 5.2 in Ref. [14] is lengthy and demanding.

Here, we implement universal expressions for scattering amplitudes and abbreviate notation using (σ′σ) ≡ $F_{\sigma'\sigma}$, etc., for amplitudes listed by Scagnoli and Lovesey, Appendix C in Ref. [15]. A similar analysis exists for polar absorption events such as E1-E2 (Appendix D in Ref. [15]), and E1-M1 where M1 is the magnetic dipole moment [26, 27]. The x-ray energy tuned to the copper $L_3$ edge (E ≈ 0.945 keV) accesses E1 (2p → 3d). Recall that the x-ray wavelength λ (Å) and energy E (keV) satisfy λ ≈ (12.4/E). An E1-E2 absorption event accesses Dirac multipoles. At the copper K edge (E ≈ 8.993 keV) the transitions are E1 (1s → 4p) and E2 (3d → 1s) using pd hybridization [28]. Set against this probable scenario, a simple atomic picture shows that at the copper K edge the E1-E2 strength is 28% of the E1-E1 strength. Magnetic properties of compounds $V_2O_3$ [29] and $TbMnO_3$ [30], for example, have been extensively investigated by resonant x-ray diffraction with the primary beam tuned to a K edge.

Quadrupoles $\langle T^2_{+1} \rangle''$ and $\langle T^2_{+2} \rangle''$, permitted by angular anisotropy in Cu atomic charge distributions, alone are revealed with a reflection vector (0, 2n + 1, 0). An E1-E1 absorption event for this special case yields diffraction amplitudes (σ′σ) = (π′π) = 0 and,

$$(\pi'\sigma) = -\alpha' \cos(\theta) [\cos(\psi) \langle T^2_{+1} \rangle'' + \sin(\psi) \langle T^2_{+2} \rangle''].  \quad (1)$$

Intensity $|(\pi'\sigma)|^2$ is two-fold periodic in the azimuthal angle ψ, namely, a sum of cos(2ψ) and sin(2ψ). The crystal axis c is normal to the plane of scattering depicted in Fig. (2). For the more general reflection vector (h, 2n + 1, 0), using an E1-E1 event we define cos(φ) = − h/[$h^2$ + (a k sin(β)/b)$^2$]$^{1/2}$. As shown in the Appendix amplitudes include spatial phase factors derived from general coordinates. In the present case the factor α = exp(i2πhx) = α′ + iα″ [11].

$$(\sigma'\sigma) = -2\ \alpha'\cos(\varphi)\sin(\psi)\ [\cos(\psi)\ \langle T^2_{+1}\rangle'' + \sin(\varphi)\sin(\psi)\ \langle T^2_{+2}\rangle''].\tag{2}$$

$$(\pi'\pi) = (\alpha''/\sqrt{2})\sin(2\theta)\ [\cos(\psi)\ \langle T^1\zeta\rangle - \sin(\varphi)\sin(\psi)\ \langle T^1\xi\rangle]$$

$$+\ \alpha'\cos(\varphi)\ [-\sin^2(\theta)\sin(2\psi)\ \langle T^2_{+1}\rangle'' + 2\sin(\varphi)\{1 - (\sin(\theta)\sin(\psi))^2\}\langle T^2_{+2}\rangle''].$$

$$(\pi'\sigma) = -(\alpha''/\sqrt{2})\ [\cos(\theta)\sin(\psi)\ \langle T^1\zeta\rangle - \{\sin(\varphi)\cos(\theta)\cos(\psi) + \cos(\varphi)\sin(\theta)\}\ \langle T^1\xi\rangle]$$

$$-\ \alpha'\ [\cos(\varphi)\sin(\theta)\cos(2\psi) + \sin(\varphi)\cos(\theta)\cos(\psi)]\ \langle T^2_{+1}\rangle''$$

$$-\ \alpha'\sin(\psi)\ [\sin(2\varphi)\sin(\theta)\cos(\psi) - \cos(2\varphi)\cos(\theta)]\ \langle T^2_{+2}\rangle''.$$

The corresponding chiral signature [12],

$$\Upsilon = \{(\sigma'\pi)^*(\sigma'\sigma) + (\pi'\pi)^*(\pi'\sigma)\}'' = 0.\tag{3}$$

The null value for $\Upsilon$ is a simple consequence of purely real diffraction amplitudes.

Our results for a space group forbidden reflection $(0, 0, 2n + 1)$ with an E1-E1 absorption event depend on $d = \cos(\beta)$ and $e = \sin(\beta)$, where the obtuse angle $\beta \approx 104.745°$ [11]. Also, there is a spatial phase factor $\gamma = \exp(i2\pi lz) = \gamma' + i\gamma''$ [11]. Unique axis b is parallel to the axis y in Fig. (2) at the start $\psi = 0$ of an azimuthal angle scan.

$$(\sigma'\sigma) = -\gamma'\sin(2\psi)\ [d\ \langle T^2_{+1}\rangle'' - e\ \langle T^2_{+2}\rangle''],\tag{4}$$

$$(\pi'\pi) = 2\sin(\theta)\cos(\psi)\ [(\gamma''/\sqrt{2})\cos(\theta)\{d\ \langle T^1\zeta\rangle + e\ \langle T^1\xi\rangle\}$$

$$-\ \gamma'\sin(\theta)\sin(\psi)\ \{d\ \langle T^2_{+1}\rangle'' - e\ \langle T^2_{+2}\rangle''\}],$$

$$(\pi'\sigma) = -(\gamma''/\sqrt{2})\cos(\theta)\sin(\psi)\ [d\ \langle T^1\zeta\rangle + e\ \langle T^1\xi\rangle] + (\gamma''\sin(\theta)/\sqrt{2})\ [e\ \langle T^1\zeta\rangle - d\ \langle T^1\xi\rangle]$$

$$-\ \gamma'\sin(\theta)\cos(2\psi)\ [d\ \langle T^2_{+1}\rangle'' - e\ \langle T^2_{+2}\rangle''] + \gamma'\cos(\theta)\sin(\psi)\ [e\ \langle T^2_{+1}\rangle'' + d\ \langle T^2_{+2}\rangle''].$$

Notably, quadrupoles alone determine $(\sigma'\sigma)$ which complements the result for $(\pi'\sigma)$ using a reflection vector $(0, 2n + 1, 0)$.

Experimental results for Dirac multipoles in $V_2O_3$ [31] and CuO [32, 33] have been published together with successful interpretations. Diffraction amplitudes for Dirac multipoles include magnetic charge $\langle G^0_0\rangle$, but the monopole does not contribute to E1-E2 amplitudes with the stipulation K = 1, 2, 3 from the triangle rule. For reflection vectors $(h, 2n + 1, 0)$ and $(0, 0, 2n + 1)$ all amplitudes are proportional to $\alpha'$ or $\gamma'$, respectively, because $\sigma_\theta \sigma_\pi = +1$ in the electronic structure factor Eq. (A1). Dirac amplitudes are complicated functions of the Bragg angle θ and azimuthal angle ψ [15]. We consider one example that demonstrates diffraction by the anapole $\langle G^1_\eta \rangle$ in $Cu_2(MoO_4)(SeO_3)$ parallel to the unique axis b. Continuing with $(h, 2n + 1, 0)$,

$$(\sigma'\sigma) \approx \alpha' \cos(\theta) [\cos(\varphi) \cos(\psi)\{- \langle G^1_\eta \rangle + (\sqrt{10}/3) \langle G^2_{+1} \rangle'\} + \sin(2\varphi) \sin(\psi) \langle G^2_{+2} \rangle']. \quad (5)$$

The result omits octupoles, for simplicity, and it vanishes for $h = 0$ ($\varphi = \pi/2$). The corresponding result for $(\pi'\pi)$ has $\cos(\theta)$ replaced by $\cos(3\theta)$ and quadrupoles of the opposite sign.

## IV. NEUTRON DIFFRACTION

Axial and Dirac multipoles $\langle t^K_Q \rangle$ and $\langle g^K_Q \rangle$, respectively, in Table I are appropriate for neutron diffraction. They depend on the magnitude of the reflection vector through radial integrals. Fig. (3) depicts their values derived from a tried and tested atomic code [34] applied to $Cu^{2+}$ ($3d^9$, $S = 1/2$).

Axial dipoles $\langle \mathbf{t}^1 \rangle$ depend on two radial integrals $\langle j_0(\kappa) \rangle$ and $\langle j_2(\kappa) \rangle$ derived from spherical Bessel functions, with $\langle j_0(0) \rangle = 1$ and $\langle j_2(0) \rangle = 0$. A useful result for $\langle \mathbf{t}^1 \rangle$ depends on spin $\langle \mathbf{S} \rangle$ and orbital angular momentum $\langle \mathbf{L} \rangle$ of copper ions [16], namely,

$$\langle \mathbf{t}^1 \rangle \approx (1/3) [2\langle \mathbf{S} \rangle \langle j_0(\kappa) \rangle + \langle \mathbf{L} \rangle (\langle j_0(\kappa) \rangle + \langle j_2(\kappa) \rangle)]. \quad (6)$$

The coefficient of orbital angular momentum $\langle \mathbf{L} \rangle$ is approximate, while $\langle \mathbf{t}^1 \rangle = (1/3) \langle 2\mathbf{S} + \mathbf{L} \rangle$ for $\kappa \to 0$ is an exact result. There are a total of four radial integrals for the atomic configuration $d^9$. Even rank multipoles are proportional to a radial integral of the same order, e.g., $\langle \mathbf{t}^2 \rangle \propto \langle j_2(\kappa) \rangle$. Such multipoles are forbidden if the electronic wave-function is created from a state with one value of the total angular momentum J. In consequence, $\langle \mathbf{t}^2 \rangle$ and $\langle \mathbf{t}^4 \rangle$ are delicate measures of J-mixing caused by the spin-orbit interaction. Such is the case for $d^9$ in an octahedral crystal field ($O_h$ symmetry) where $J = 3/2$ and $J = 5/2$ define the ground state [35]. Results in Fig. 3 show that $\langle j_2(\kappa) \rangle$ peaks around $\kappa \approx 6$ Å$^{-1}$, which is indicative of the κ range to be covered in measuring magnetization distributions [36].

The Dirac dipole $\langle \mathbf{d} \rangle$ in neutron diffraction depends on three radial integrals displayed in Fig. (3). We use [16],

$$\langle \mathbf{d} \rangle = (1/2) [ i(g_1) \langle \mathbf{n} \rangle + 3 (h_1) \langle \mathbf{S} \times \mathbf{n} \rangle - (j_0) \langle \mathbf{\Omega} \rangle]. \quad (7)$$

Radial integrals $(g_1)$ and $(j_0)$ diverge in the forward direction of scattering ($\kappa \to 0$), and $(h_1)$ is also the κ-dependence of the polar spin quadrupole observed in neutron diffraction from high-$T_c$ compounds Hg1201 and YBCO [37, 38]. Dipoles $\langle \mathbf{S} \times \mathbf{n} \rangle$ and $\langle \mathbf{\Omega} \rangle = [\langle \mathbf{L} \times \mathbf{n} \rangle - \langle \mathbf{n} \times \mathbf{L} \rangle ]$ are spin and orbital anapoles.

The amplitude of magnetic neutron diffraction $\langle \mathbf{Q}_\perp \rangle = [\langle \mathbf{Q} \rangle - \mathbf{e}(\mathbf{e} \cdot \langle \mathbf{Q} \rangle)]$ yields an intensity $|\langle \mathbf{Q}_\perp \rangle|^2 = \{|\langle \mathbf{Q} \rangle|^2 - |(\mathbf{e} \cdot \langle \mathbf{Q} \rangle)|^2\}$, where the unit vector $\mathbf{e} = \mathbf{\kappa}/\kappa$ [16]. With our phase convention in the electronic structure factor Eq. (A1), which incorporates $\langle O^K_Q \rangle^* = (-1)^Q \langle O^K_{-Q} \rangle$ for the complex conjugate of a multipole, axial and polar $\langle \mathbf{Q} \rangle$ are purely imaginary, and nuclear structure factors are purely real. In consequence, nuclear and magnetic intensities are in quadrature. Our magnetic amplitudes appear as purely real, however, because we elect to omit the factor $i = \sqrt{(-1)}$. A polarized neutron diffraction signal $\Delta = \{\mathbf{P} \cdot \langle \mathbf{Q}_\perp \rangle\}$, where $\mathbf{P}$ is polarization of the primary neutrons. A spin-flip intensity SF is a measure of the magnetic content of a Bragg spot, and SF $= \{|\langle \mathbf{Q}_\perp \rangle|^2 - \Delta^2\}$ given $\mathbf{P} \cdot \mathbf{P} = 1$ and $(\langle \mathbf{Q}_\perp \rangle^* \times \langle \mathbf{Q}_\perp \rangle) = 0$. Notably, $\Delta = 0$ for parallel $\mathbf{P}$ and $\mathbf{e}$.

With $\mathbf{e} = (e_\xi, e_\eta, 0)$ and odd $k$ magnetic neutron amplitudes include dipoles parallel to axes $\xi$ and $\zeta$, as in the corresponding x-ray diffraction amplitudes Eq (2). Axial amplitudes up to and including quadrupoles are,

$$\langle Q_\xi \rangle \approx \alpha'' [(3/2) \langle t^1_\xi \rangle - e_\eta^2 \langle t^2_{+1} \rangle''], \quad \langle Q_\eta \rangle \approx \alpha'' e_\xi e_\eta \langle t^2_{+1} \rangle'',$$

$$\langle Q_\zeta \rangle \approx \alpha'' [(3/2) \langle t^1_\zeta \rangle + (e_\xi^2 - e_\eta^2) \langle t^2_{+2} \rangle'']. \tag{8}$$

SF $= \langle Q_\zeta \rangle^2$ using in-plane $\mathbf{P} = (-e_\eta, e_\xi, 0)$ to achieve $(\mathbf{e} \cdot \mathbf{P}) = 0$, and SF $= (\mathbf{e} \times \langle \mathbf{Q} \rangle)_\zeta^2$ on using $\mathbf{P} = (0, 0, 1)$. In the foregoing results, $e_\xi = h/[h^2 + (a\, k\, \sin(\beta)/b)^2]^{1/2}$ and $e_\eta = (a\, k\, \sin(\beta)/b)/[h^2 + (a\, k\, \sin(\beta)/b)^2]^{1/2}$.

For reflections indexed by $(0, 0, 2n + 1)$ we use a notation $\Phi = [d\, \langle t^2_{+2} \rangle'' + e\, \langle t^2_{+1} \rangle'']$ with, as before, $d = \cos(\beta)$, $e = \sin(\beta)$, and find,

$$\langle Q_\xi \rangle \approx \gamma'' [(3/2) \langle t^1_\xi \rangle + e\, \Phi], \quad \langle Q_\eta \rangle \approx 0, \quad \langle Q_\zeta \rangle \approx \gamma'' [(3/2) \langle t^1_\zeta \rangle + d\, \Phi]. \tag{9}$$

SF $= 0$ for polarization $\mathbf{P} = (e, 0, d)$, and SF $= |\langle \mathbf{Q}_\perp \rangle|^2$ for $\mathbf{P} = (0, 1, 0)$ with $(\mathbf{e} \cdot \mathbf{P}) = 0$ in both cases.

Parity-odd additions to $\langle \mathbf{Q}_\perp \rangle$ are identified in our results by a superscript $^{(-)}$. For a reflection vector $\mathbf{e} = (e_\xi, e_\eta, 0)$ and odd $k$,

$$\langle Q_{\perp\xi} \rangle^{(-)} \approx 2\, \alpha'\, e_\xi\, e_\eta^2\, \langle g^2_{+2} \rangle', \quad \langle Q_{\perp\eta} \rangle^{(-)} \approx -2\, \alpha'\, e_\xi^2\, e_\eta\, \langle g^2_{+2} \rangle',$$

$$\langle Q_{\perp\zeta} \rangle^{(-)} \approx \alpha'\, e_\xi\, [\langle g^1_\eta \rangle - \langle g^2_{+1} \rangle']. \tag{10}$$

Bragg spots indexed by $(0, 0, 2n + 1)$ with $\mathbf{e} = (-d, 0, e)$ are created by Dirac amplitudes,

$$\langle Q_{\perp\xi} \rangle^{(-)} \approx -\gamma'\, e\, [\langle g^1_\eta \rangle + d\, e\, \{\langle g^2_{+2} \rangle' - \sqrt{(3/2)} \langle g^2_0 \rangle\}], \quad \langle Q_{\perp\eta} \rangle^{(-)} \approx 0, \tag{11}$$

$$\langle Q_{\perp\zeta}\rangle^{(-)} \approx -\gamma' d\,[\langle g^1_{\eta}\rangle + d\,e\,\{\langle g^2_{+2}\rangle' - \sqrt{(3/2)}\,\langle g^2_0\rangle\} - d\,\{d^2 - e^2\}\langle g^2_{+1}\rangle'].$$

Working exclusively with dipoles, the total magnetic amplitude is,

$$\langle \mathbf{Q}_\perp\rangle + \langle \mathbf{Q}_\perp\rangle^{(-)} \approx (\{\gamma''\,e^2\,(3\langle t^1\xi\rangle/2) - \gamma'\,e\,\langle g^1_{\eta}\rangle\},\ 0,\ \{\gamma''\,d^2\,(3\langle t^1\zeta\rangle/2) - \gamma'\,d\,\langle g^1_{\eta}\rangle\}). \quad (12)$$

Here and in the foregoing results, $\gamma = \exp(i2\pi l/z) = \gamma' + i\,\gamma''$ [11].

## V. CONCLUSIONS

Our survey of symmetry informed magnetic Bragg diffraction patterns for the monoclinic transition metal dioxomolybdenum selenite $Cu_2(MoO_4)(SeO_3)$ is based on the magnetic structure $P2_1'/c$ (No. 14.77, magnetic crystal class $2'/m$) proposed by Piyawongwatthana *et al*. [11]. The authors concluded that at a low temperature axial dipoles derived from $Cu^{2+}$ (S = 1/2) align antiferromagnetically along the crystal axis c with weak noncollinearity. Two independent copper ions are in Wyckoff general positions 4e that are devoid of symmetry. The crystal structure is centrosymmetric, and ferromagnetism and the piezomagnetic effect are forbidden in $2'/m$. Commensurate antiferromagnetic order with a propagation vector $= (0, 0, 0)$ is established below a temperature $\approx 23$ K. Parity (P) and time (T) symmetries are paired in the magnetic crystal class, i.e., it is (PT)-symmetric as a consequence of anti-inversion ($\bar{1}'$) in $2'/m$. Piyawongwatthana, *et al*. confirmed the $2'/m$ symmetry of the tensor for a linear magnetoelectric effect [11].

Results for neutron and x-ray diffraction patterns given here include Dirac multipoles that are polar and magnetic. Likely, the Dirac monopole and dipole, or anapole, are the most familiar. Neutron and x-ray scattering amplitudes are derived from an electronic structure factor that incorporates all symmetry present in $P2_1'/c$ and Wyckoff general positions 4e. With this in hand, universal expressions for the amplitudes in terms of electronic multipoles with discrete symmetries are at our disposal.

**ACKNOWLEDGEMENT** Dr Ph. Sainctavit gave advice on copper K-edge absorption.

## APPENDIX: ELECTRONIC STRUCTURE FACTOR AND MULTIPOLES

Our structure factor [14, 15, 16],

$$\Psi^K_Q(4e) = [\exp(i\mathbf{\kappa}\cdot\mathbf{d})\,\langle O^K_Q\rangle_{\mathbf{d}}],$$

$$= [\alpha\gamma + \sigma_\pi\,\sigma_\theta\,(\alpha\gamma)^*]\,[\langle O^K_Q\rangle + \sigma_\pi\,(-1)^{K+Q}\,(-1)^{k+l}\langle O^K_{-Q}\rangle], \quad (A1)$$

delineates a Bragg diffraction pattern for a reflection vector $\mathbf{\kappa}$ defined by integer Miller indices (h, k, l) and generic electronic multipoles $\langle O^K\rangle$ of rank K. The sum is over positions $\mathbf{d}$ used by

Cu ions in $P2_1'/c$ (No. 14.77). They are Wyckoff general positions 4e for Cu1 and Cu2 in $Cu_2(MoO_4)(SeO_3)$, and devoid of symmetry. The electronic structure factor encompasses all symmetries of sites 4e in magnetic symmetry $P2_1'/c$. Spatial phase factors in Eq. (A1) are $\alpha = \exp(i2\pi h x)$ and $\gamma = \exp(i2\pi l z)$. Estimates of general coordinates (x, 0, z) for Cu ions appear in Table II of Ref. [11]. Space group (basis) forbidden reflection conditions are identified in Eq. (A1) by considering its value for even K and Q = 0, and parity signature $\sigma_\pi = +1$, i.e., the structure factor for nuclear scattering. The result is $[1 + (-1)^{k+l}] = 0$ for Miller indices with odd $k + l$. Bulk magnetic properties are defined by $\Psi^K_Q(4e)$ evaluated for K = 1, $\sigma_\theta = -1$ and $h = k = l = 0$. The corresponding result $\Psi^K_Q(4e) \propto [1 - \sigma_\pi] \langle O^1_Q \rangle'$ confirms an antiferromagnetic structure of axial dipoles and anapoles, $\langle \mathbf{g}^1 \rangle$ or $\langle \mathbf{G}^1 \rangle$ in Table I, in the plane normal to the unique axis b.

In more detail, the result in Eq. (A1) requires information about the relevant Wyckoff positions found in the Bilbao table MWYCKPOS for the magnetic symmetry of interest [13]. Wyckoff positions are related by operations listed in the table MGENPOS [13]. Taken together, the two tables provide all information required to evaluate Eq. (A1) and, thereafter, all x-ray and neutron diffraction amplitudes. The result depends on the discreet symmetries of the multipole $\langle O^K_Q \rangle$, where projections Q are in the range $-K \leq Q \leq Q$. Angular brackets in $\langle O^K_Q \rangle$ denote the time-average, or expectation value, of the enclosed spherical tensor operator. Table I contains the parity signature $\sigma_\pi = +1$ ($-1$) for axial (polar) multipoles. The time signature $\sigma_\theta = -1$ for magnetic neutron and Dirac multipoles, and $\sigma_\theta = (-1)^K$ for axial multipoles observed in x-ray diffraction enhanced by a parity-even absorption event, e.g., E1-E1.

Axial multipoles $\langle \mathbf{T}^K \rangle$ use reduced matrix elements Eq. (73) in Ref. [14]. Therein a unit tensor that is Eq. (2.8) in Ref. [16] for a single hole ($3d^9$) [35]. Appendix A in Ref. [39] includes expressions for some Dirac multipoles. Multipoles inferred from experimental data can be confronted with estimates using a particular atomic wave-function for the resonant ion [33, 40, 41], or simulations of electronic structure [42-45].

Cartesian and spherical components Q = 0, $\pm 1$ of a vector $\mathbf{n} = (\xi, \eta, \zeta)$ are related by $\xi = (n_{-1} - n_{+1})/\sqrt{2}$, $\eta = i(n_{-1} + n_{+1})/\sqrt{2}$, $\zeta = n_0$. A complex conjugate of a multipole is defined as $\langle O^K_Q \rangle^* = (-1)^Q \langle O^K_{-Q} \rangle$, meaning the diagonal multipole $\langle O^K_0 \rangle$ is purely real. The phase convention for real and imaginary parts labelled by single and double primes is $\langle O^K_Q \rangle = [\langle O^K_Q \rangle' + i \langle O^K_Q \rangle'']$. Whereupon, Cartesian dipoles are $\langle O^1_\xi \rangle = -\sqrt{2} \langle O^1_{+1} \rangle'$ and $\langle O^1_\eta \rangle = -\sqrt{2} \langle O^1_{+1} \rangle''$.

Table I contains a summary of multipoles that we use. Axial (parity even) multipoles possess a time signature $(-1)^K$. They can contribute to diffraction enhanced by E1-E1 or E2-E2 absorption events. All multipoles are functions of the quantum numbers that define the core state of photo-ejected electrons. The dependence on quantum numbers manifests itself in so-called sum rules that relate $\langle \mathbf{O}^K \rangle$ measured at $L_2$ and $L_3$ edges, for example. Sum rules for the axial dipole $\langle \mathbf{T}^1 \rangle$ present the orbital angular momentum $\langle \mathbf{L} \rangle$ in the valence state [25]. Dirac atomic multipoles $\langle \mathbf{G}^K \rangle$ are polar (parity odd) and magnetic (time odd) [14, 33]. They are permitted in a magnetic material when the resonant ion occupies an acentric site. Detection of Dirac multipoles require a probe with matching attributes, of course, which are found in x-ray diffraction enhanced by E1-E2 or E1-M1 parity-odd absorption events.

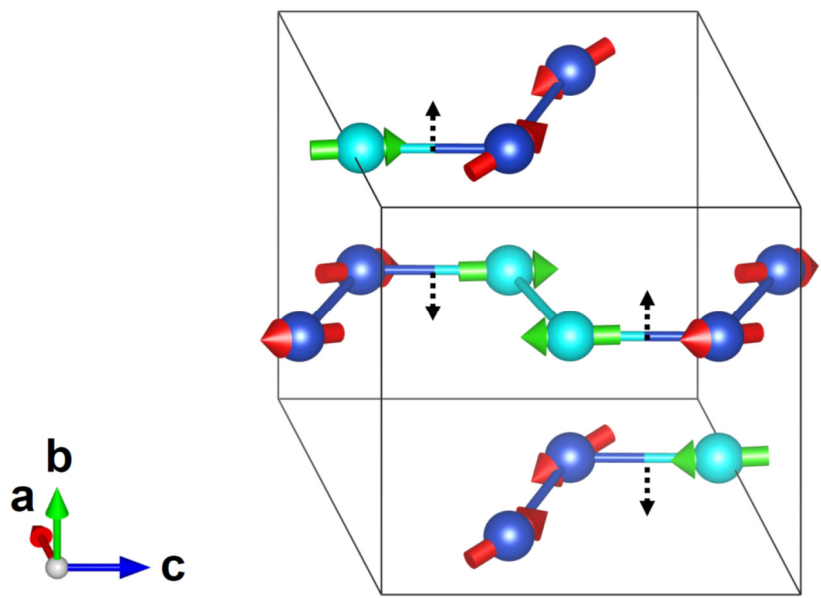

**FIG. 1**. Axial dipoles in $Cu_2(MoO_4)(SeO_3)$ using magnetic symmetry $P2_1'/c$ (No. 14.77 [13]). They are in Wyckoff general positions 4e for Cu1 (red arrows) and Cu2 (green). Black dashed arrows indicate the components of the Dzyaloshinskii-Moriya interaction along the unique axis b. The cartoon appears as figure 7b in reference [11], and it is reproduced here by courtesy of the authors.

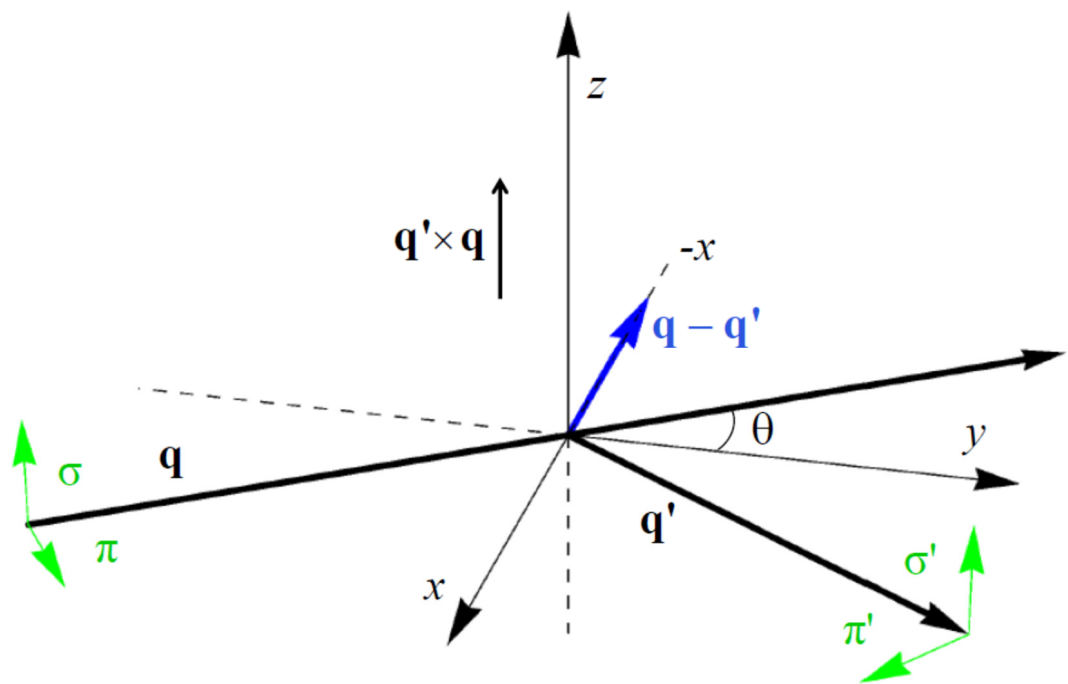

**FIG. 2**. Primary (σ, π) and secondary (σ', π') states of polarization. Corresponding wavevectors **q** and **q'** subtend an angle 2θ. The Bragg condition for diffraction is met when **q** − **q'** coincides with a reflection vector (*h*, *k*, *l*) of the monoclinic reciprocal lattice. Crystal vectors that define

local axes (ξ, η, ζ) and the depicted Cartesian (x, y, z) coincide in the nominal setting of the crystal.

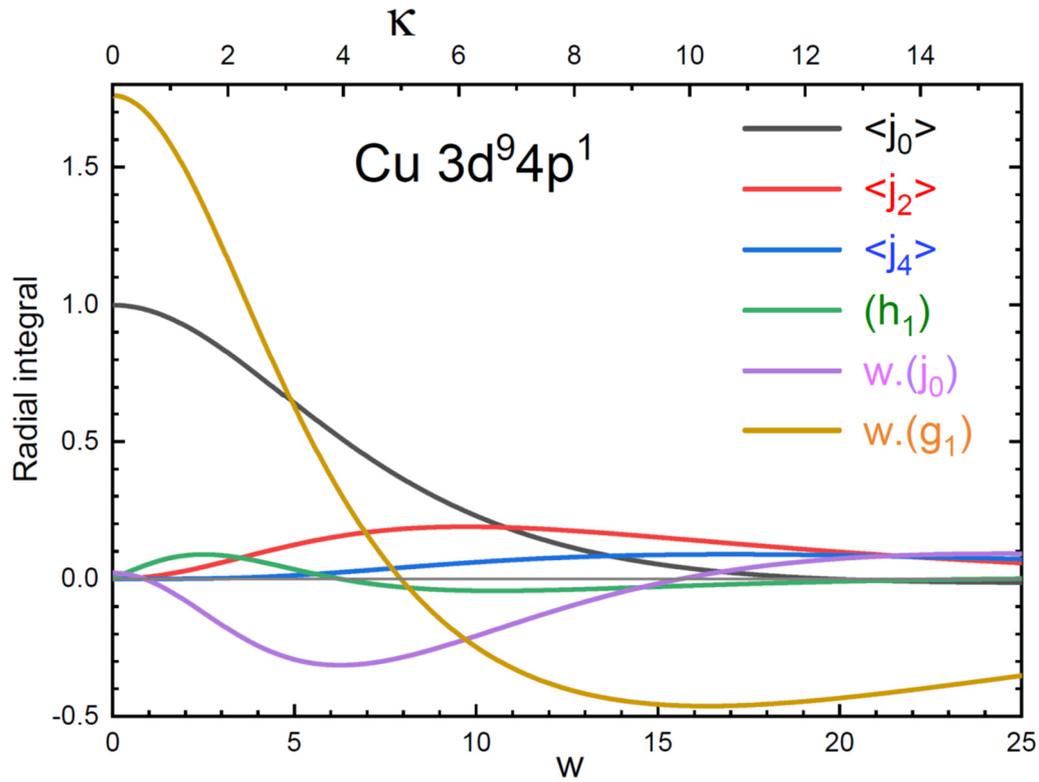

**FIG. 3**. Radial integrals for $Cu^{2+}$ ($3d^9$, spin = 1/2) derived from Cowan's atomic code [34] appearing in Eqs. (6) and (7). The magnitude of a reflection vector $\kappa = (4\pi/\lambda) \sin(\theta)$, where $\lambda$ (Å) is the neutron wavelength and $\theta$ is the Bragg angle in Fig. 2. A second variable w is rendered dimensionless by the Bohr radius $a_o$ in $w = 3\, a_o\, \kappa$.

**TABLE I.** A generic multipole $\langle O^K{}_Q \rangle$ has integer rank K and (2K + 1) projections Q. Parity ($\sigma_\pi$) and time ($\sigma_\theta$) signatures = ±1, e.g., $\langle t^K{}_Q \rangle$ for magnetic neutron diffraction is parity-even ($\sigma_\pi = +1$) and time-odd ($\sigma_\theta = -1$). Copper ions in $Cu_2(MoO_4)(SeO_3)$ use Wyckoff positions that have no symmetry. In consequence, all Q are permitted. Also, Dirac multipoles $\langle g^K{}_Q \rangle$ and $\langle G^K{}_Q \rangle$ with $\sigma_\theta \sigma_\pi = +1$ are allowed. The phase convention for real and imaginary parts labelled by single and double primes is $\langle O^K{}_Q \rangle = [\langle O^K{}_Q \rangle' + i\langle O^K{}_Q \rangle'']$.

| Signature | $\sigma_\pi$ | $\sigma_\theta$ |
|---|---|---|
| *Neutrons* | | |
| $\langle t^K{}_Q \rangle$ | +1 | −1 |
| $\langle g^K{}_Q \rangle$ | −1 | −1 |
| | | |
| *Photons* | | |
| $\langle T^K{}_Q \rangle$ | +1 | $(-1)^K$ |
| $\langle G^K{}_Q \rangle$ | −1 | −1 |